\documentclass[twocolumn, secnumarabic, amssymb, amsmath, nobibnotes, aps, prx, superscriptaddress]{revtex4-2}

\usepackage{graphicx,comment}
\usepackage{float}
\usepackage{dcolumn}
\usepackage{bm}
\usepackage{braket}
\usepackage[dvipsnames]{xcolor}
\usepackage[colorlinks=true]{hyperref}
\definecolor{darkblue}{RGB}{33,33,137}
\definecolor{darkred}{RGB}{193,23,23}
\hypersetup{
    colorlinks=true,       
    linkcolor=blue,     
    citecolor=blue,    
    urlcolor=blue      
} 
\usepackage{natbib}
\usepackage[caption=false]{subfig}
\usepackage{color,soul}
\usepackage{tikz,fp}
\usepackage{tikz-cd}
\usepackage{environ}
\usepackage{upgreek} 
\usepackage{relsize}
\usepackage{wasysym}

\usepackage[normalem]{ulem}

\definecolor{darkblue}{rgb}{0.1,0.2,0.6}
\definecolor{crimson}{RGB}{164,16,52}

\begin{document}

\title{Lattice Defects in Rydberg Atom Arrays}
\author{Hanteng Wang}
\affiliation{
Institute for Advanced Study, Tsinghua University, Beijing 100084, China
}

\author{Chengshu Li}
\affiliation{
Institute for Advanced Study, Tsinghua University, Beijing 100084, China
}

\author{Xingyu Li}
\affiliation{
Institute for Advanced Study, Tsinghua University, Beijing 100084, China
}

\author{Yingfei Gu}
\affiliation{
Institute for Advanced Study, Tsinghua University, Beijing 100084, China
}

\author{Shang Liu}
\email{sliu.phys@gmail.com}
\affiliation{
Kavli Institute for Theoretical Physics, University of California, Santa Barbara, California 93106, USA
}
\affiliation{Department of Physics, California Institute of Technology, Pasadena, California 91125, USA}

\date{February 11, 2025}

\begin{abstract}
Rydberg atom arrays have become a key platform for studying quantum many-body systems. In these setups, defects arise naturally due to various imperfections and can significantly modify the theoretical predictions compared to an ideal model. Here, we investigate the impact of geometric defects in the simplest situation -- a one-dimensional Rydberg atom array, both at and away from its emergent Ising criticality. In the presence of defects, we demonstrate that relevant physical quantities can be extracted from one-point correlation functions. At the critical point, we show that different types of kinks yield distinct outcomes corresponding to their respective spatial-internal symmetries: site-centered kinks can effectively break the array at the kink position regardless of the kink angle, while bond-centered kinks lead to interesting intermediate-coupling fixed points. In the latter case, due to a special renormalization group flow trajectory, the whole system can appear ordered if the system is not large enough. Additionally, away from criticality, the bond-centered kink induces a localization-delocalization transition of the domain wall, characteristic of quantum wetting. These findings highlight the utility of kinks as experimental probes and stress the importance of controlling defects so that experimental observations remain faithful to the pristine model.

\end{abstract}

\maketitle

\section{Introduction}
\label{sec:Intro}
Understanding the emergence of universal behavior is a central theme in quantum many-body physics. Quantum criticality \cite{CardyBook,SachdevBook}, characterized by scale invariance, is a fundamental arena for investigating complex interactions and phase transitions in both natural and engineered systems, often through the lens of conformal field theory (CFT) \cite{DiFrancescoCFTBook}. Recent advances have highlighted the potential of simulating these critical phenomena using controllable, high-precision quantum platforms \cite{zhang2017observation,harris2018phase}.

Among these platforms, Rydberg atom arrays have emerged as a powerful and versatile tool for investigating quantum many-body physics \cite{bernien2017probing,browaeys20,Ebadi2021,Scholl2021,Semeghini2021,fang2024probing,manovitz2024quantum} and pushing forward quantum computing capabilities \cite{Saffman2010,morgado2021quantum,Bluvstein2022,evered2023high,scholl2023erasure,ma2023high,Singh23,bluvstein2024logical}. These experiments have enabled the study of diverse quantum phases across different spatial dimensions. Although long-range order in strictly one-dimensional systems with short-range interaction is typically precluded by thermal fluctuations \cite{mermin1966absence}, improvements in optical-tweezer techniques have allowed the realization of intriguing phases, such as the $\mathbb{Z}_2$ Ising phase \cite{fendley2004competing,slagle2021microscopic}, $\mathbb{Z}_3$ phases \cite{samajdar2018numerical,whitsitt2018quantum}, and the floating phase \cite{weimer2010two,chepiga2019floating,rader2019floating,maceira2022conformal} to be observed. Quasi-one-dimensional structures, such as two-leg ladders, further enrich the landscape of possible phases \cite{Li2024,Chepiga2024}.

Transitions between distinct phases often involve nontrivial critical phenomena that uncover deep theoretical structures and insights, such as confinement-deconfinement transitions in gauge theories \cite{surace2020lattice,cheng2023variational,cheng2024emergent} and manifestations of space-time supersymmetry \cite{Li2024}. 
In addition to equilibrium studies, Rydberg atom arrays provide a platform to explore non-equilibrium dynamics, including the Kibble-Zurek mechanism for probing critical behavior \cite{keesling2019quantum,chepiga2021kibble} and Floquet dynamics \cite{feldmeier2024quantum,köylüoğlu2024floquet,liu2024supersymmetry}.

Despite significant advances with Rydberg atom arrays, experimental imperfections and defects remain an inevitable challenge. For example, due to the discrete pixelation constraints of spatial light modulators, using circular geometry \cite{fang2024probing} introduces greater uncertainty in the atomic positions compared to a straight-line arrangement. Hence, one may reduce the on-site positional uncertainty in atomic positions by rearranging atoms into a square configuration, but at the cost of creating local geometric defects. These deformations can significantly impact the system’s behavior and lead to outcomes that deviate from ideal predictions. In such square configurations, unavoidable kinks can have pronounced effects both at and away from criticality. It is well established that, near critical points, defects can induce substantial, nontrivial modifications, as exemplified by the famous Kondo effect \cite{affleck1995conformal}. As a result, defect CFT has been developed to address these phenomena \cite{cardy1991bulk,billo2016defects,makabe2017defects,ge2024defect}, and it continues to be an active research area with numerous open questions \cite{andrei2020boundary}.

In this work, we investigate the impact of geometric defects within the $1/R^6$ Rydberg blockade model, focusing on the simplest system -- a one-dimensional array of atoms, which hosts an Ising-type quantum phase transition upon adjusting the detuning parameter. 
By analyzing the role of kink deformations, we aim to elucidate their influence on the low-energy physics and provide practical insights for experimental implementations. 
We have analyzed multiple scenarios and found the following three settings particularly interesting. 
\begin{enumerate}
    \item \textbf{Site-centered kink at criticality}: The slogan is ``kink = cut''; such a defect effectively cuts open the system at the kink location. 
    
    \item \textbf{Bond-centered kink at criticality}:
The kink generates a novel defect renormalization group (RG) flow that has a interesting effect at intermediate energy scale: The $\mathbb{Z}_2$ order parameter value increases as the system size enlarges, thus it seems that the kink induces ordering throughout the whole system. Eventually in the infrared, the kink will not affect the bulk criticality, but will be described by an intermediate-coupling fixed point. This means that the kink effect is neither vanishing nor segmenting the system as in the above situation. 
 
    \item \textbf{Bond-centered kink in the ordered phase}: A domain wall localization-delocalization transition, also known as a quantum wetting transition, can be generated by tuning the kink angle. 
\end{enumerate}
It is rather remarkable that such rich phenomena all emerge from a very simple setup. These findings are supported by both analytical approaches and density matrix renormalization group (DMRG) simulations. 

The rest of this paper is organized as follows. We define our models more concretely in Sec~\ref{sec:Models}. In Sec~\ref{sec:critical}, we analyze in detail the roles of kinks at the quantum critical point. In Sec~\ref{sec:localization}, we explore the behavior of kinks as the detuning shifts into the ordered phase. Additional technical details and numerical data are provided in the appendices.

\section{Models, defects and boundaries} 
\label{sec:Models}
 \subsection{Pristine model} 
We consider a one-dimensional array of $N$ atoms, where each atom can occupy either the ground state $|g_i\rangle = |\Circle\rangle$ or a Rydberg excited state $|r_i\rangle = |\CIRCLE\rangle$. The presence of a Rydberg excitation at site $i$ is quantified by $\hat{n}_i = |r_i\rangle \langle r_i| = (1 + \hat{Z}_i)/2$. The transition between the ground and Rydberg excited states is facilitated by a Rabi coupling, $\hat{X}_i = |r_i\rangle \langle g_i| + \text{h.c.}$, characterized by the Rabi frequency $\Omega$. This setup also includes the repulsive van der Waals interactions between atoms at sites $i$ and $j$, introducing the Rydberg blockade effect, which depends on the occupancy of Rydberg excitations. This leads to a tractable experimental model with the Hamiltonian
\begin{equation} \label{R6}
  H =  \frac{\Omega}{2} \sum_{i=1}^{N} \hat{X}_i - \Delta \sum_{i=1}^{N} \hat{n}_i + \sum_{i \neq j} V_{ij} \hat{n}_i \hat{n}_j ,
\end{equation}
where $\Delta$ is the detuning of the driving field, and the two-body Rydberg interaction between atoms $i$ and $j$ is given by $V_{ij} = \Omega\left( R_b/R_{ij} \right)^6$.

Our focus is on the nearest neighbor Rydberg blockade regime, where a translation symmetry-breaking transition is observed with increasing detuning $\Delta$. We design the system such that nearest neighbors, separated by the lattice constant $a$, are within the Rydberg blockade radius $R_b$, while next-nearest neighbors are not, i.e., $1 < R_b/a < 2$. Hence, simultaneous excitation of adjacent atoms is not favored by a strong energy penalty $V_1/\Omega = (R_b/a)^6 > 1$, while the next-nearest neighbor energy penalty $V_2/\Omega = [R_b/(2a)]^6 < 1$ is weak. This setup can also be approximately described by the PXP model \cite{turner2018weak, turner2018quantum,moudgalya2022quantum, yao2022quantum} in which the nearest-neighbor Rydberg interaction is set to infinity.
Ising criticality occurs at $\Delta = \Delta^{\mathbb{Z}_2}_c$ (approximately 1.1, for $R_b/a = 5/4$, with $\Omega=1$ set throughout the paper), where the system transitions from a disordered phase, represented by $|\cdots\Circle\Circle\Circle\Circle\cdots\rangle$, to a phase for $\Delta > \Delta^{\mathbb{Z}_2}_c$ with two degenerate symmetry-breaking states exhibiting density wave order, either $|\cdots\Circle\CIRCLE\Circle\CIRCLE\cdots\rangle$ or $|\cdots\CIRCLE\Circle\CIRCLE\Circle\cdots\rangle$, with standard periodic boundary conditions with even $N$. In the continuum description of this phase transition \cite{yao2022quantum,slagle2021microscopic}, the single-site translation symmetry effectively becomes a $\mathbb{Z}_2$ symmetry. Hence, we will subsequently also refer to this critical point as a $\mathbb{Z}_2$ symmetry breaking transition.

 \begin{figure}[t]
  \centering
  \includegraphics[width=0.47\textwidth]{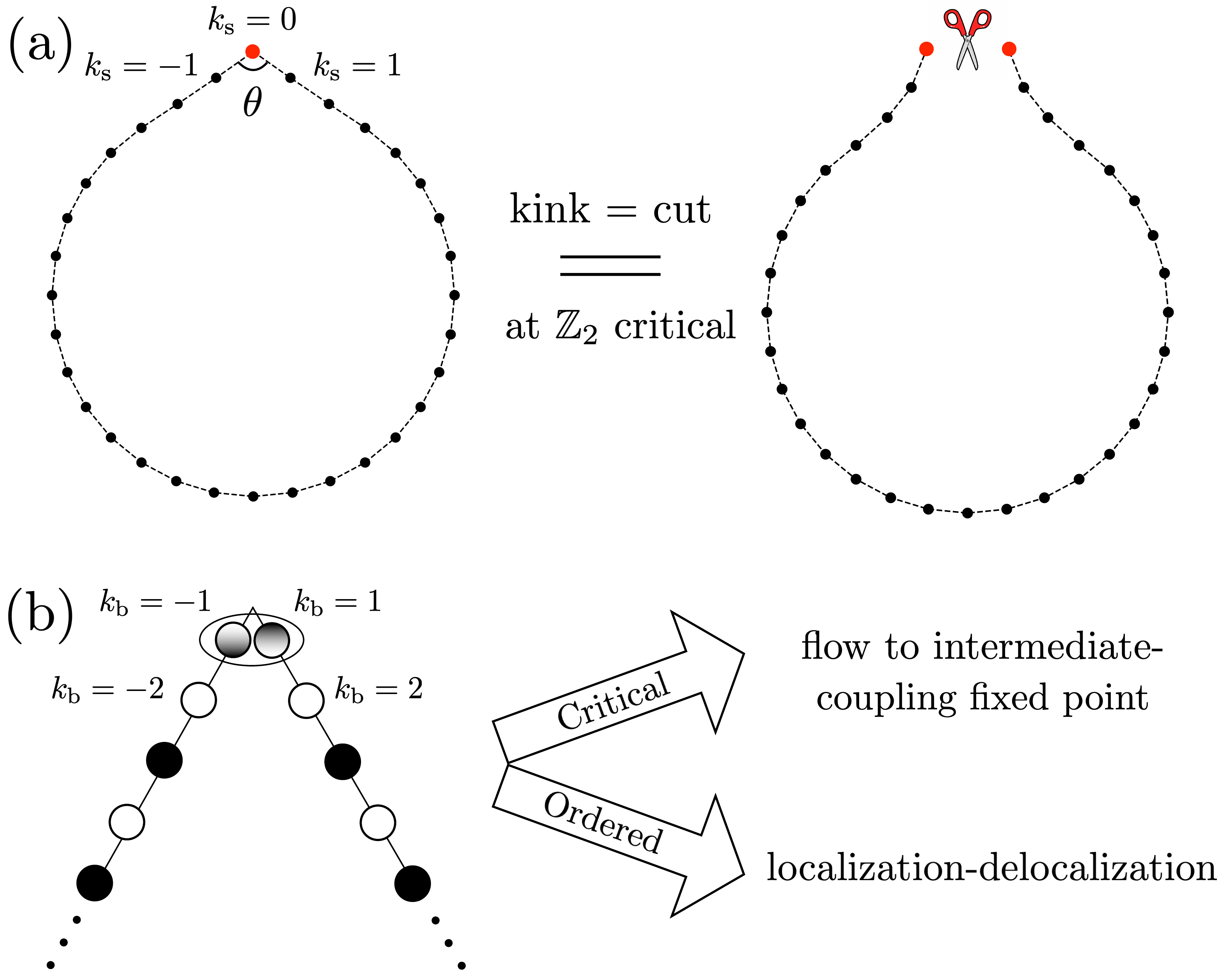}
  \caption{(a) Site-centered kink deformation (Eq.~\eqref{eq:site_kink}) increases the interaction strength between its two neighbors, effectively breaking the chain at the $\mathbb{Z}_2$ critical point. (b) Bond-centered kink deformations  (Eq.~\eqref{eq:bond-center_kink}) preserves an extra symmetry compared to (a), driving the system to intermediate-coupling fixed points at $\mathbb{Z}_2$ criticality. Additionally, a localization-delocalization transition emerges in the ordered phase.}
  \label{fig:kink_cut}
\end{figure}

In this subsection, we explore the impact of local defects in Rydberg atom arrays, which are commonly encountered in experimental setups. We focus on two types of defects: (i) site-centered (s) type kinks and (ii) bond-centered (b) type kinks.

Firstly, for the site-centered type kink, we define the position of the center atom as $k_\text{s}=0$. Atoms to the left are labeled as $k_\text{s} = -1, -2, -3, \dots$, and to the right as $k_\text{s} = 1, 2, 3, \dots$. Positioning $k_\text{s}=0$ as the pivot center, we introduce a kink in the array at an angle $\theta$, which modifies the geometric arrangement as depicted in Fig.~\ref{fig:kink_cut}(a). The distance between the atoms at $k_\text{s}=\pm1$ is adjusted by the kink angle to $2a\sin(\theta/2)$. This affects the interaction between the atoms at $k_\text{s}=\pm1$ sites, which is described by
\begin{equation}\label{eq:site_kink}
\begin{aligned}
&\hat{\mathcal{V}}_\text{s}(\theta) = V_{\text{s},1} \, \hat{n}_{k_\text{s}=-1}\hat{n}_{k_\text{s}=1};\\
&V_{\text{s},1}(\theta)=\left[\frac{R_b/a}{2\sin(\theta/2)}\right]^6,
\end{aligned}
\end{equation}
where $V_{\text{s},1}(\theta)$ represents the modified nearest neighbor Rydberg repulsion due to the site-centered kink.

Secondly, the bond-center kink involves bending the chain between two atoms as Fig.~\ref{fig:kink_cut}(b). We label atoms with respect to the bond center from the left as $k_\text{b} = -1, -2, -3, \dots$, and from the right as $k_\text{b} = 1, 2, 3, \dots$. The nearest $k_\text{b} = \pm1$ sites now have their distances altered to $a\sin(\theta/2)$, and the next-nearest neighbor distance for $k_\text{b} = \pm1, \mp2$ is modified to $a\sqrt{(5-3\cos(\theta))/2}$. Thus, the interactions among these sites are given by
\begin{equation}\label{eq:bond-center_kink}
\begin{aligned}
\hat{\mathcal{V}}_\text{b}(\theta) =  {}& V_{\text{b},1}\, \hat{n}_{k_\text{b}=-1}\hat{n}_{k_\text{b}=1} 
\\&+  V_{\text{b},2}\,( \hat{n}_{k_\text{b}=1}\hat{n}_{k_\text{b}=-2}+\hat{n}_{k_\text{b}=-1}\hat{n}_{k_\text{b}=2});\\
V_{\text{b},1}(\theta) = {}& \left[\frac{R_b/a}{\sin(\theta/2)}\right]^6,\, V_{\text{b},2}(\theta)=\left[\frac{R_b/a}{\sqrt{(5-3\cos\theta)/2}}\right]^6,
\end{aligned}
\end{equation}
where $V_{\text{b},1}(\theta)$ and $V_{\text{b},2}(\theta)$ represent the nearest and next nearest neighbor Rydberg repulsion modified by the bond-centered kink.

\subsection{Boundaries of arrays} 
The low-energy physics in a Rydberg array with nearest neighbor blockade is heavily influenced by the parity of the site number $N$ and the type of boundary conditions applied, even when the boundary is distant from defect positions. Therefore, it is crucial to specify the parity of $N$ (even or odd) and the type of boundary (closed or open) when analyzing these systems.

We overview how open vs closed boundaries interact with different parities of $N$, which leaves one configuration frustrated while the other not. For both odd $N$ with periodic boundaries and even $N$ with open boundaries, domain wall structures emerge in the low-energy sector, marking a significant difference from a simple $\mathbb{Z}_2$ symmetry-breaking scenario with just two degenerate states. For instance, consider an array with $N=6$ and open boundaries: if the blockade energy scale $V_1$ is substantially high, then in the ordered phase, states such as $|\Circle\CIRCLE\Circle\CIRCLE\Circle\CIRCLE\rangle$, $|\CIRCLE\Circle\Circle\CIRCLE\Circle\CIRCLE\rangle$, $|\CIRCLE\Circle\CIRCLE\Circle\Circle\CIRCLE\rangle$, and $|\CIRCLE\Circle\CIRCLE\Circle\CIRCLE\Circle\rangle$ constitute the lowest energy band, all sharing the same classical energy. Unlike the periodic case with even $N$, where the two degenerate $\mathbb{Z}_2$ symmetry-breaking phases are separated by many $\hat{X}_i$ flips and the non-local cat state is not stable, in an open chain, adjacent domain wall states are connected by just two $\hat{X}_i$ flips. As a result, they can easily mix due to quantum fluctuations induced by $\hat{X}_i$, lowering the energy to form a ground state that is an almost equal superposition of all these states. Similarly, for odd $N$ with closed boundaries, the same frustration effect is observed, revealing an intriguing similarity between
\begin{itemize}
  \setlength{\itemindent}{-0em}
  \item[] \textit{closed odd/even array} $\sim$ \textit{open even/odd array}.
\end{itemize}

In the following sections, we will thoroughly investigate the 1D Rydberg array in the nearest neighbor blockade regime, examining different types of kinks in conjunction with various boundary conditions. Section \ref{sec:critical} will discuss how kinks affect the system at the $\mathbb{Z}_2$ critical point. Section \ref{sec:localization} will explore the kink effect deep within the ordered phase.

\section{Kink effects at $\mathbb{Z}_2$ criticality}
\label{sec:critical}
In this section, we explore how various types of kinks influence the low energy physics at the $\mathbb{Z}_2$ critical point. First, we will discuss some preliminaries that help us understand the low-energy physics of the model without the influence of kinks.

\subsection{Continuum theory of $\mathbb{Z}_2$ universality}
In the pristine Rydberg array within the constraint $1 < R_b/a < 2$, the system transitions from a disordered state to a $\mathbb{Z}_2$ ordered state across a critical point. This critical point shares the same universality class of Ising criticality with the canonical microscopic example of the transverse field Ising model and PXP model. In the low energy sector of our pristine Rydberg model, the local $\mathbb{Z}_2$ order parameter $Z_i - Z_{i+1}$, which condenses on the ordered side of the transition, corresponds to a spin field $\phi(x)$. The continuum spatial coordinate $x$ and the lattice site index $i$ are related by $x/L=i/N$, where $L\propto N$ is the total system size of the continuum theory \footnote{Conventionally, we choose $L/N$ such that the ``speed of light'' is equal to $1$. }. 
Thus in path integral formalism, this local order parameter $\phi(\tau,x)$ can be used for a coarse-grained $1+1$-dimensional action within the Landau-Ginzburg paradigm
\begin{equation}
S_0(\Delta) = \int d\tau d x \left[ \frac{1}{2} (\partial\phi)^2+\frac{r_0(\Delta)}{2}\phi^2+\frac{u_0(\Delta)}{4}\phi^4\right],
\end{equation}
where the detuning $\Delta$ governs the transition. At the critical point, $\phi$ exhibits a power-law decay of the correlation function
\begin{equation}
\langle \phi(x)\phi(x^\prime)\rangle \sim |x-x^\prime|^{-2\Delta_\phi}
\end{equation}
with (anomalous) scaling dimension $\Delta_\phi = 1/8$.

To connect the microscopic description to its low energy effective theory at the critical point, one can note that the polarization operator $\hat{Z}_i$ in a critical Rydberg chain can be related to the spin field in its operator version through an order-by-order expansion \footnote{In the standard notations of Ising CFT, $\phi\propto\sigma$ and $:\phi^2:\propto\varepsilon$, where $::$ refers to normal ordering. }
\begin{equation}
\label{eq:order_by_order_expansion}
\hat{Z}_i \sim c_0+c_1(-1)^i\hat{\phi}+ c_2 \hat{\phi}^2 + c_{(1,1)} \partial_x\hat{\phi}+ \cdots.
\end{equation}
Consequently, identifying the properties, such as the scaling dimension of the $\hat{\phi}$ operator, can provide insights into observable effects in experiments around the low energy critical point.

In the subsequent parts of this section, we aim to introduce defects to the pristine array and examine their response. As we will demonstrate, the two types of defects $\hat{\mathcal{V}}_\text{s}$ and $\hat{\mathcal{V}}_\text{b}$ exhibit distinctly different effects. We will discuss these from an RG perspective, supported by numerical checks using DMRG methods.

\subsection{Destiny of different kinks}
In this part, we explore the impact of kinks based on general arguments. To begin with, it is essential to determine the most appropriate quantity for measurement in the experiment. Guided by the continuous theory discussed earlier, $\hat Z_i$ encapsulates information about the order parameter scaling dimension $\Delta_\phi$. Therefore, measuring the two-point correlation of $\hat{Z}_i$ appears to be a promising approach. In the pristine model, measuring the two-point correlation is indeed necessary to detect $\Delta_\phi$. However, in the presence of kinks, one-point functions suffice to measure the scaling dimension, since the internal $\mathbb{Z}_2$ symmetry is explicitly broken.

To motivate why a two-point correlator might be effectively replaced by a one-point measurement, consider a system slightly detuned from criticality where first-order quantum mechanical perturbation theory applies. We define the quantity
\begin{equation}
\label{eq:delta_Z}
\delta Z_i(\theta) = \langle \text{GS}(\theta) |\hat{Z}_i|\text{GS}(\theta)\rangle - \langle \text{GS}(\pi) |\hat{Z}_i|\text{GS}(\pi)\rangle,
\end{equation}
which quantifies how a kink modifies the on-site diagonal $Z$ operator measurement, where $|\text{GS}(\theta)\rangle$ denotes the ground state with kink angle $\theta$. Assuming $|\text{GS}(\theta)\rangle$ can be perturbatively derived from $|\text{GS}(\pi)\rangle$ at the first order of the kink operator $\hat{\mathcal{V}}$, and given the system is gapped away from criticality, we find $\delta Z_i (\theta) \propto \langle \text{GS}(\pi)|\hat{\mathcal{V}} \hat{Z}_i|\text{GS}(\pi)\rangle_\text{con}$. This relationship suggests that by measuring a one-point correlator, information about the two-point correlation can be inferred. However, this approach is only valid when $\Delta$ is away from its critical value $\Delta_c^{\mathbb{Z}_2}$, where correlators decay exponentially. This simple perturbative approach holds as detailed in Appendix \ref{app:beyond_criticality}; but at the critical point, where the spectral gap closes and correlations decay algebraically, this formula is quantitatively inaccurate. Nonetheless, the principle of encoding two-point information of a pristine model within a one-point measurement with a defect remains valid and beneficial. 

Returning to the critical regime and its continuum description, based on the above suggestion, we proceed to calculate  
$\langle \phi \rangle_{\text{kink}}=\int\mathcal D[\phi]e^{-S}\phi/\int\mathcal D[\phi]e^{-S}$, 
where $S = S_0 + \delta S_\text{s}$ or $S = S_0 + \delta S_\text{b}$, representing the action modified by two types of kinks, $\hat{\mathcal{V}}_\text{s}$ or $\hat{\mathcal{V}}_\text{b}$ on top of the pristine action $S_0(\Delta_c^{\mathbb{Z}_2})$ at the critical point. We suggest the modification part of the action as
\begin{align}
\delta S_\text{s} &= \int d\tau d x \,h(\theta) \phi \cdot \delta(x) \label{eq:action_site} \\
\delta S_\text{b} &= \int d\tau d x \,[\lambda(\theta)\partial_x\phi +\mu(\theta)\phi^2] \cdot \delta(x) \label{eq:action_bond-center}
\end{align}
Here, $h(\theta)$ and $(\lambda(\theta), \mu(\theta))$ are the effective couplings of site-centered and bond-centered type kinks at angle $\theta$, respectively, and $\delta(x)$ indicates that the kink is located at $x=0$. 
Using symmetry arguments, we may validate the above actions induced by the kinks as follows. 
A kink is invariant under the spatial inversion with respect to the kink center. It is known from previous works \cite{yao2022quantum,slagle2021microscopic} that the site-centered lattice inversion maps $\phi(x)$ to $\phi(-x)$ in the continuum, i.e. the action on the order parameter is trivial except for the coordinate change $x\mapsto -x$. This can be intuitively understood by observing that the low-energy classical configuration is invariant under
\begin{align*}
|\Circle\CIRCLE\Circle\hspace{-5.4pt}\raisebox{0pt}{\textbar}\hspace{2.5pt}\CIRCLE\Circle\rangle \xrightarrow{x\mapsto-x} |\Circle\CIRCLE\Circle\hspace{-5.4pt}\raisebox{0pt}{\textbar}\hspace{2.5pt}\CIRCLE\Circle\rangle. 
\end{align*}
This allows $\phi$ to be the most relevant perturbation for a site-centered kink. In contrast, the bond-centered lattice inversion maps $\phi(x)$ to $-\phi(-x)$. Intuitively, this is because achieving invariance in a classical low-energy configuration requires two steps:
\begin{align*}
|\Circle\CIRCLE\Circle\raisebox{0pt}{\textbar}\CIRCLE\Circle\CIRCLE\rangle \xrightarrow{x\mapsto-x} |\CIRCLE\Circle\CIRCLE\raisebox{0pt}{\textbar}\Circle\CIRCLE\Circle\rangle \xrightarrow{\phi\mapsto-\phi}|\Circle\CIRCLE\Circle\raisebox{0pt}{\textbar}\CIRCLE\Circle\CIRCLE\rangle. 
\end{align*}
Consequently, $\phi$ is no longer an allowed perturbation for a bond-centered kink, while higher-order terms $\phi^2$ and $\partial_x\phi$ are permitted.

These actions, still at the UV limit with degrees of freedom coarsely grained by $\phi$, raise the question: How does the low-energy (IR) physics change compared to the non-kinked pristine case when we RG flow these actions? For the site-centered kink, we understand that the perturbation $\phi$ is relevant, with a scaling dimension of $1/8$. This relevant perturbation will flow the system to another fixed point, where at the kink location, $|h| \rightarrow \infty$, and hence $\phi$ at the kink position is fully polarized. 
For an intuitive understanding, we start from a comparatively strong kink effect scenario.
In the Rydberg lattice language, the boundary value of $|Z_0-Z_1|$ and $|Z_{-1}-Z_0|$ flows to $\mathcal{O}(1)$, acting as a Rydberg-favoring pinning field at the kink. Consequently, at the critical point, the kink effectively breaks the correlation between the left and right parts of $k = 0$, mimicking a cut, creating open boundaries and effectively adding one site, as depicted in Fig.~\ref{fig:kink_cut}(a).
For weaker kink effects, with a large enough system size, we expect the same consequence due to RG flow. 
Thus, we predict:
\begin{itemize}
  \setlength{\itemindent}{-1em}
  \item[] \textit{closed odd/even array with site-centered kink} \\$\simeq$ \textit{open even/odd array}.
\end{itemize}
This means the closed even/odd array with a site-centered kink eventually flows to an open odd/even array at low energy, for a sufficiently large system size $N$, or for a small finite $N$ with a small enough angle $\theta$ (which already has a strong kink effect by hand at the UV level). For more details, in Appendix \ref{app:Rydberg_vs_Ising}, we perform a microscopic level correspondence between the Rydberg array and the Ising description. On the other hand, the bond-centered perturbation has two terms: $\phi^2$ which by itself is an \emph{exactly} marginal defect perturbation with a scaling dimension of $1$, and $\partial_x\phi$ which is weakly irrelevant with a scaling dimension of $1 + 1/8$. When both terms are small, the low-energy physics does not deviate significantly from the pristine fixed point, and correlation functions can be accurately computed by field-theoretic perturbation theory. 
However, when the perturbations get larger, there is actually an exotic defect RG flow with interesting finite-size phenomena as we will discuss later.

Hence for these two types of kinks, we argued that they will have very different destinies in the thermodynamic limit. For more details, we will discuss them in the following sections for site-centered kink (Sec.~\ref{sec:site_kink}) and bond-centered kink (Sec.~\ref{sec:bond-center_kink}).

\subsection{Site-centered kink: ``kink = cut''}
\label{sec:site_kink}
In this subsection, we explore the site-centered kink scenario, performing numerical checks via DMRG to corroborate our ``kink = cut'' slogan. Specifically, we measure the one-point function, revealing the scaling dimension $\Delta_\phi = 1/8$, described as $\delta Z_i \sim (-1)^i\phi$. The $\delta Z_i$ values are determined using Eq.~(\ref{eq:delta_Z}) or almost equivalently calculated as $\delta Z_i = \langle \hat{Z}_i - \hat{Z}_{i+1} \rangle/2$, after obtaining the ground state wavefunction at $\Delta_c^{\mathbb{Z}_2}$ with kink at $k_\text{s}=0$. Both methods provide equivalent insights into the field $\phi$ and the scaling dimension $\Delta_\phi$ due to the order-by-order expansion Eq.~(\ref{eq:order_by_order_expansion}). We predict that once the site-centered kink effect is significant, the lowest energy state evolves towards a fixed point, characteristic of systems with open boundaries. Consequently,  for long distance, $\delta Z_i$ should conform to the conformal scaling expected in finite systems \cite{Burkhardt1993IsingBCFT,slagle2021microscopic}, as
\begin{equation}
\label{eq:Zi_curve}
\delta Z_i = 
\begin{cases} 
\mathlarger{\frac{(-1)^i\lambda}{\left[N\sin\left(\pi x/L\right)\right]^{\Delta_\phi}}}, & \text{for even }N \\[25pt]
\mathlarger{\frac{(-1)^i\lambda \cos\left(\pi x/L\right)}{\left[N\sin\left(\pi x/L\right)\right]^{\Delta_\phi}}}, & \text{for odd }N
\end{cases}
\end{equation}
where $\lambda$ is a constant beyond the scope of CFT. 
Here we relabel site $i$ to $N+i$ for $-N/2<i < 0$ in the data of $\delta Z_i$ from kink set-up, ensuring a smooth function $\delta Z_i$ with respect to $x$ defined over $0 \leq x/L \leq 1$.

\begin{figure}[t]
  \centering
  \includegraphics[width=0.4\textwidth]{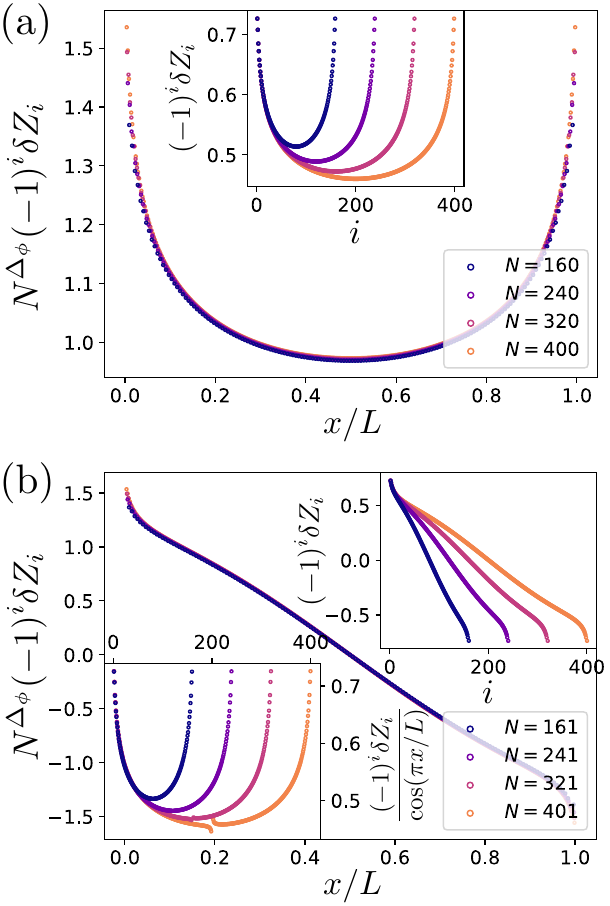}
  \caption{Plot of $(-1)^i\delta Z_i = (-1)^i\langle \hat{Z}_i - \hat{Z}_{i+1} \rangle/2$  vs sites $i$ with site-centered kink at angle $\theta=60^\circ$, for (a) even $N$ and (b) odd $N$ cases (insets). The main figures depict data collapsed onto a universal curve by setting $N^{\Delta_\phi} (-1)^i\delta Z_i$ vs $x/L$.}
  \label{fig:even_odd}
\end{figure}

\begin{figure}[t]
  \centering
  \includegraphics[width=0.4\textwidth]{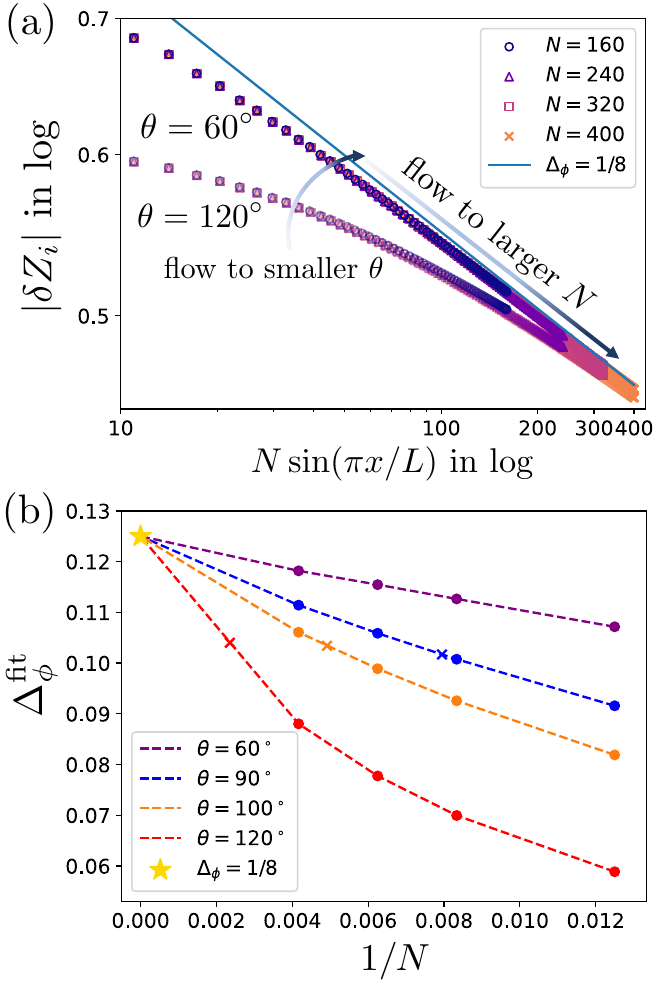}
  \caption{(a) Log-log plot of $|\delta Z_i|$ vs $N\sin(\pi x)$, for $\theta=60^\circ$ and $\theta=120^\circ$ (fader colors). (b) Fitted scaling dimension $\Delta_{\phi}^{\text{fit}}$ with increasing system size $N$ for site-centered kink angles $\theta=60^\circ, 90^\circ, 100^\circ, 120^\circ$. Crosses on each curve label the estimated limiting system size that begins to obey CFT prediction, set by $g = 60$.}
  \label{fig:flow}
\end{figure}

Figure \ref{fig:even_odd} illustrates how $\delta Z_i$ changes with respect to $x$ for a kink angle $\theta = 60^\circ$. This angle corresponds to a strong coupling of $h(\theta)$ in action (\ref{eq:action_site}), prompting us to expect the data to conform to Eq.~(\ref{eq:Zi_curve}). Indeed, in the main plot of Fig.~\ref{fig:even_odd}, both the even and odd $N$ cases show data collapse when the $y$-axis is scaled as $N^{1/8} \delta Z_i$. Moreover, for the odd case, normalizing the data by $\delta Z_i/\cos(\pi x/L)$ reverts it to the form expected for the even case, as anticipated. The presence of the $\cos(\pi x/L)$ factor can qualitatively be understood as distinguishing between periodic and anti-periodic boundary conditions, where the latter must rotate from positive to negative sign of $\delta Z_i$ upon returning to the kink position. The even/odd dependency in Eq.~(\ref{eq:Zi_curve}) is exactly opposite to that observed in the open boundary Rydberg chain in Ref.~\cite{slagle2021microscopic}, which leads us to the statement that a \textit{closed odd/even array} $\simeq$ \textit{open even/odd array}. In addition to system size scaling $\sim N^{-\Delta_\phi}$, we also explicitly check the spatial relationship $\sim x^{-\Delta_\phi}$, as shown in Fig.~\ref{fig:flow}(a). We plot $|\delta Z_i|$ against $N\sin(\pi x/L)$ on a log-log scale for $\theta=60^\circ$ and $120^\circ$. With a fixed $N$, increasing the detection scale, $x$, progressively aligns the curve with a straight decay with a slope of $1/8$, indicative of $\Delta_\phi=1/8$. Thus, the data presented so far affirm the validity of our ``kink = cut'' statement.

Yet, experimental settings often raise questions such as: What defect strength is necessary for the CFT predictions of Eq.~(\ref{eq:Zi_curve}) to be accurate for a given system size $N$? Conversely, if the defect is minor, how large must the system size be for the kink effect to be significant?  Previous discussions have been qualitative, focusing on the ultimate destination of the RG fixed point without detailing the ``kink $\rightarrow$ cut'' flowing process. In standard RG flow process, one frequently sits at thermodynamic limit and decreases the energy scale. However, we may translate the standard RG point of view to how the ground state converges towards CFT predictions as the system size increases.

To examine the numerical manifestation of the RG flow, we aim to characterize how closely the numerical data of $\delta Z_i$ compare to Eq.~(\ref{eq:Zi_curve}) across different system sizes and kink angles. Let us reexamine Fig.~\ref{fig:flow}(a) and extract its RG information: as $N$ increases, the data points start to converge to the same curve and align more and more closely with the $1/8$ slope line. This indicates two pieces of information: larger values of $x$ (but need a constraint with $x/L<1/2$) probe lower energy sectors, approaching CFT predictions; and increasing $N$ similarly drives the system towards CFT results, suggesting that the thermodynamic destiny of a strong kink is equivalent to an open array. When plotting data for $\theta=120^\circ$ marked with crosses, the weaker defect effect necessitates a larger scale to probe and converge to the RG fixed point. As another side of the same coin of the RG picture, at a fixed position of $x$, stronger kinks (i.e., lower $\theta$ values) push $\delta Z_i$ closer to CFT predictions. The dual effects of RG flow, increasing $N$ (or $x$) and decreasing $\theta$, are indicated by arrows with a color gradient. This qualitative picture of RG flow is complemented by quantitative analysis in Fig.~\ref{fig:flow}(b). We fit the data at positions far from the kink to determine the slope of $\log|\delta Z_i|$ vs $\log[N\sin(\pi x/L)]$, denoted as $\Delta_{\phi}^{\text{fit}}$. At infinitely large $N$ and very small $\theta$, this slope is expected to approach $1/8$, marked by the yellow star in Fig.~\ref{fig:flow}(b). For finite $N$ and intermediate $\theta$ values, Fig.~\ref{fig:flow}(b) illustrates how $\Delta_{\phi}^{\text{fit}}$ converges towards its destination of $1/8$. 

Furthermore, we develop a simple qualitative criterion to estimate how large $N$ is needed for a given kink to approximate the CFT prediction. Considering the ground state $|\text{GS}(\theta)\rangle$, we approximate it from the no-kink ground state $|\text{GS}(\pi)\rangle$ using a perturbative approach: $|\text{GS}(\theta)\rangle \approx |\text{GS}(\pi)\rangle + \frac{\hat{\mathcal{V}}\text{s}}{E_1-E_0} |\text{GS}(\pi)\rangle$, where $E_0$ and $E_1$ are the unperturbed ground and first excited states, separated by $\mathcal{O}(1/N)$ near the critical point. The matrix element between these two states is proportional to the local interaction strength $V_{\text{s},1}(\theta)$ defined in Eq.~(\ref{eq:site_kink}), validating the perturbation if $g(\theta, N) \equiv V_{\text{s},1}(\theta) N \ll 1$. Conversely, if $g \gg 1$, the system behavior should closely align with CFT predictions. Setting this dimensionless parameter to a value significantly greater than one yields a critical system size, larger than which the system approximates CFT predictions well. These critical values are marked with crosses in Fig.~\ref{fig:flow}(b). For example, for $g = 60$, this analysis results in a scaling dimension around 0.105, offering a practical approach for experimental setups without extensive calculations.

\subsection{Bond-centered kink}
\label{sec:bond-center_kink}
We turn our attention to the bond-centered kink case. The lattice model is now invariant under a spatial inversion with respect to the kink location. As we mentioned previously, such a bond-centered lattice inversion acts in the continuum field theory as $\phi(x)\mapsto -\phi(-x)$. Based on this symmetry constraint, we write down the defect action as 
\begin{align}
    \delta S_\text{b} &= \int d\tau d x \,[\mu\phi^2 + \lambda\partial_x\phi + \cdots] \cdot \delta(x), 
\end{align}
where $\cdots$ represents less relevant operators. The $\phi^2$-term on its own is known to be an \emph{exactly} marginal perturbation \footnote{This may be understood by observing that in the dual fermion description of Ising CFT, $\phi^2$ is mapped to a fermion bilinear term. For a thorough study of the $\phi^2$ defect in Ising CFT, see Ref.~\cite{Oshikawa1997IsingDCFT}. }. In other words, if only this term is turned on, then there is no defect RG flow, and each distinct value of $\mu$ defines a different defect fixed point. The $\partial_x\phi$ term on its own is a weakly irrelevant defect perturbation. 

\begin{figure}[t]
  \centering
  \includegraphics[width=0.4\textwidth]{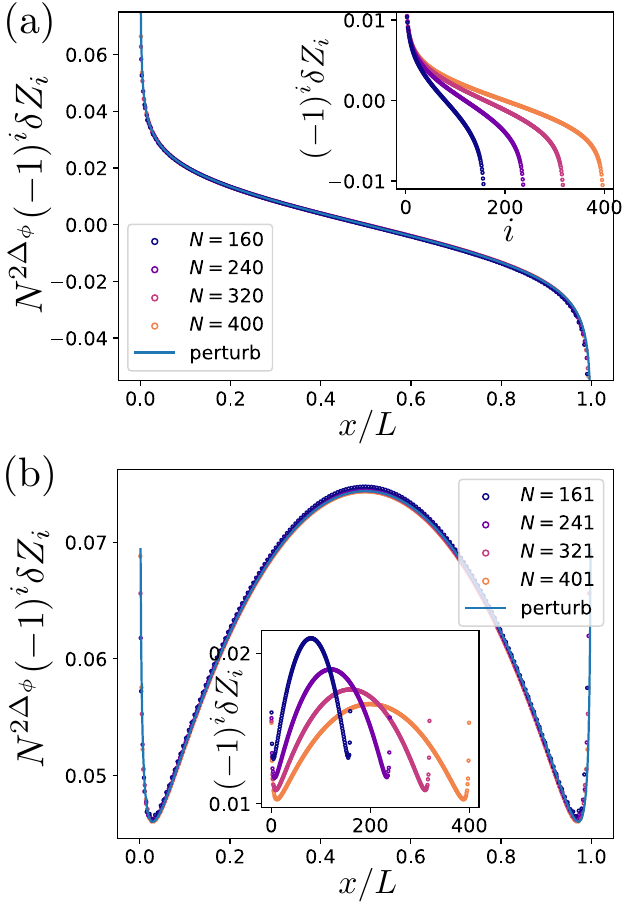}
  \caption{Plot of $\delta Z_i$ vs even sites $i$ with kink angle $\theta=140^\circ$ positioned at $k_\text{b}=0$, for (a) even $N$ and (b) odd $N$ cases. The main figures depict data collapsed onto a curve by setting $N^{2\Delta_\phi} \cdot (-1)^i \delta Z$ vs $x$, where $\Delta_\phi=1/8$.}
  \label{fig:even_odd_bond-center}
\end{figure}

\begin{figure}[t]
  \centering
  \includegraphics[width=0.4\textwidth]{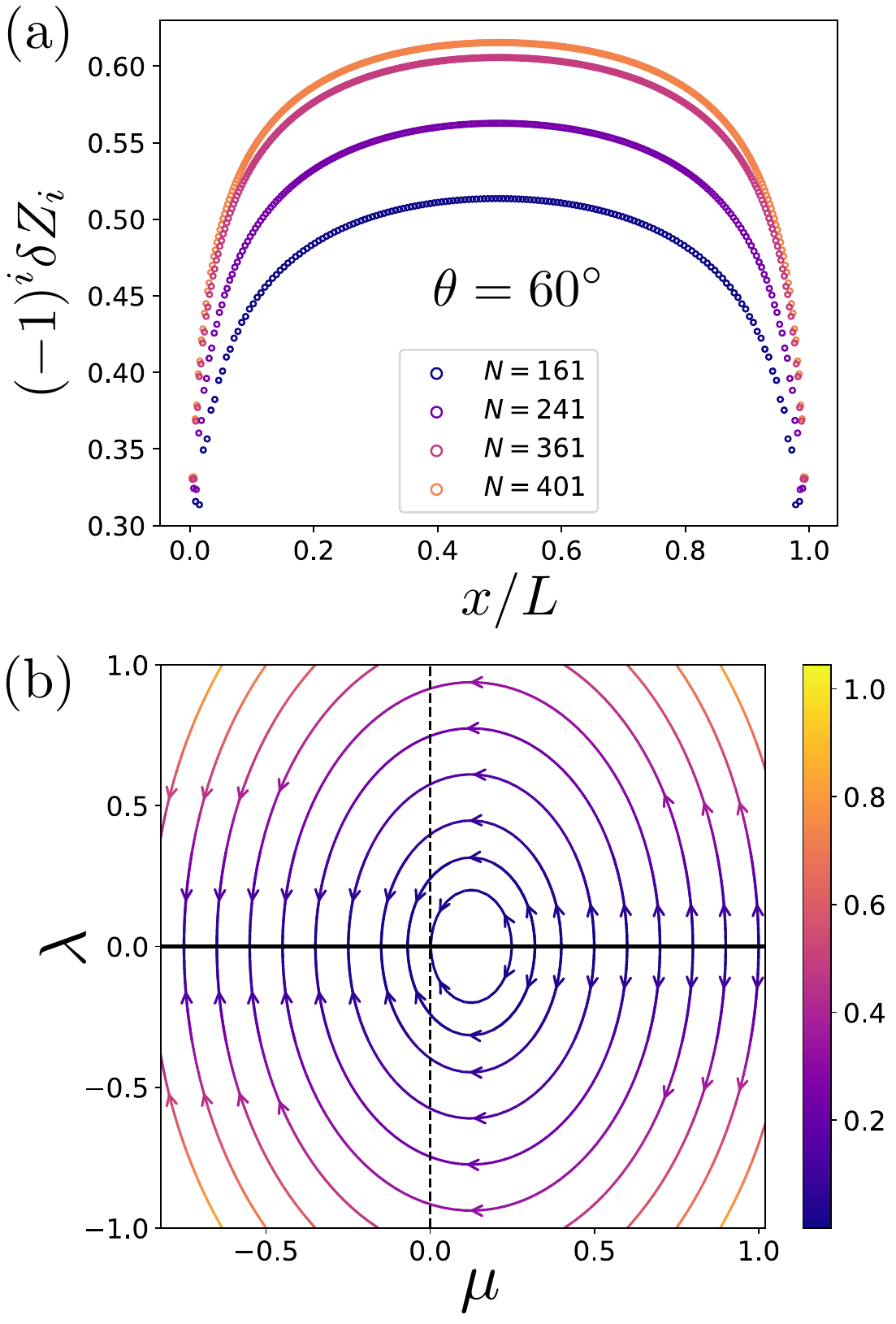}
  \caption{(a) For $\theta=60^\circ$, with increasing system size up to $N\simeq400$, $(-1)^i \delta Z_i$ is still increasing. (b) 1-loop perturbtive RG. The color bar refers to the flow speed.}
  \label{fig:RG_bond-center}
\end{figure}

When all the couplings in the defect action are very small, we may compute correlation functions using first-order perturbation theory in the path integral. For example, the one-point function of $\phi(x)$ is approximately given by
\begin{align}
    \braket{\phi(x)}\approx \braket{\phi(x)}_0+\braket{(-\delta S_{\rm b})\mathcal \phi(x)}_{0,\text{con}}, 
    \label{eq:BondCenteredKinkVevPerturbative}
\end{align}
where $\braket{\cdot}_0$ stands for the expectation value in the unperturbed theory, and the subscript ``con'' stands for the connected correlation function. We emphasize that the above statement would not be valid if there were a relevant or marginally relevant perturbation which effectively became large at low energy. Equation (\ref{eq:BondCenteredKinkVevPerturbative}) implies that 
\begin{align}
\braket{\phi(x)}\approx 
N^{-1/4} f\left({2\pi i}/{N}\right)\quad(x/L=i/N), 
\end{align}
where the function $f(\cdot)$ can be computed using known results of Ising CFT correlators \cite{DiFrancesco1987IsingCFT,DiFrancescoCFTBook}
For completeness, let us present the explicit form of $f(\cdot)$ below. 
When $N$ is even, which corresponds to the periodic boundary condition $\phi(x+L)=\phi(x)$ in the continuum, we have found
\begin{align}
    f_{\rm even}(y)&\propto -\int_{-\infty}^\infty d\tau\,\partial_y \frac{e^{\tau/8}}{r(\tau,y)^{1/4}}, \\
    \text{where}\quad r(\tau,y)&:=\sqrt{(e^\tau-\cos y)^2+\sin^2y}. 
\end{align}
When $N$ is odd, which corresponds to the antiperiodic boundary condition $\phi(x+L)=-\phi(x)$ in the continuum, we have found 
\begin{align}
    f_{\rm odd}(y)\propto -\int_{-\infty}^\infty d\tau\,\partial_y\sqrt{ \frac{e^{\tau/2}+e^{-\tau/2}}{r(\tau,y)^{1/2}}-e^{-\tau/2}r(\tau,y)^{1/2} }. 
\end{align}
Recall that $\hat\phi$ is the leading nontrivial operator in the expansion of $(-1)^i\hat Z_i$. We thus expect the above perturbative result to well describe $(-1)^i\delta Z_i$, provided that the kink perturbation is weak, or the kink angle is close to $\pi$, which is necessary to justify the first-order perturbation theory. This is verified by our numerical simulation. Figure \ref{fig:even_odd_bond-center} demonstrates the case of numerical calculation of $\langle \hat{Z}_i \rangle$ with both even and odd $N$ at small kink strength ($\theta=140^\circ$), which verifies the scaling of $N^{1/4}$ and the form of the function $f(2\pi i/N)$ explicitly.

As we further decrease the kink angle (increase the strength of the kink perturbation), we start observing deviations from the previous perturbative results. Most remarkably, we have observed that for sufficiently small kink angles, the overall value of $(-1)^i\delta Z_i$ decreases much slower or even increases as $N$ gets larger, at least within the range of system sizes accessible to us; see Fig.~\ref{fig:RG_bond-center}(a) for kink angle $60^\circ$ with up to $N=401$. In the following, we will provide a possible qualitative explanation of this exotic phenomenon by examining the defect RG at 1-loop order. 

The defect RG equations can be derived using conformal perturbation theory; see e.g. Refs.\,\cite{Gaberdiel2009ConformalPertTheory,CardyBook}. Denote by $\tilde\mu$ and $\tilde\lambda$ some properly normalized dimensionless versions of $\mu$ and $\lambda$, respectively. We have found the following at 1-loop order. 
\begin{align}
    \frac{d\tilde\mu}{dl}&=-\frac{3}{8}\tilde\lambda^2+\cdots \\
    \frac{d\tilde\lambda}{dl}&=-\frac{1}{8}\tilde\lambda+\tilde\lambda\tilde\mu+\cdots
\end{align}
The corresponding RG flow diagram is given in Fig.~\ref{fig:RG_bond-center}. Evidently, $(\tilde\mu,\tilde\lambda)=(1/8,0)$ is a special point -- the center of the circular RG flow. Given the smallness of $1/8$, we expect the 1-loop perturbative RG equations to be sufficient to capture the physics near this point. From the RG equations and the flow diagram, we see that when $\tilde\mu$ and $\tilde\lambda$ are both very small, the defect RG is essentially captured by the tree-level terms: $\tilde\mu$ barely changes, and $\tilde\lambda$ flows to zero with the rate $-1/8$. This is the regime where we can use low-order perturbation theory to compute correlation functions. In contrast, when the defect couplings get larger, especially when $\tilde\mu>1/8$, the defect RG flow is drastically different: the coupling $\tilde\lambda$ can increase for a while before eventually dropping down. Note that the $\lambda$ term is the most relevant defect perturbation that breaks the Ising $\mathbb{Z}_2$ symmetry $\phi\mapsto-\phi$ and thus contributes to $\braket{\phi}$. The above defect RG flow then suggests that $\braket{\phi}$ can increase with $N$ for a range of system sizes before eventually dropping down. This serves as a possible explanation of the numerically observed increment of $\braket{\phi}$. 

\section{Away from Criticality: Localization-Delocalization Transition}
\label{sec:localization}

In this section, we investigate the effects of a kink away from the critical point within the ordered phase. For small kink effects, given the exponential decay of connected correlators in the ordered phase, little difference is expected compared to the pristine case. However, with sufficient kink strength, a new phase may emerge, leading to a quantum localization-delocalization transition.

\subsection{Strong bond-centered kink favoring localization}

In the following discussion, we focus on the scenario where the kink is of the bond-centered type. As we will see, site-centered kinks do not exhibit similar effects due to energetic considerations. The effects discussed here may only be observable in even open arrays or odd closed arrays, since in these cases the system has a non-trivial classical, i.e. $\Omega=0$, ground-state manifold in its pristine state.

We first review the ground-state manifold and structure of the ordered phase for an open chain with an even number of sites, $N$. We adopt a bond-centered notation to label the atomic positions, dividing the chain into two equal halves, each containing $N/2$ atoms on either side of the central bond, as depicted in Fig.~\ref{fig:kink_cut}(b). In the ordered phase, where the blockade energy scale $V_1 = (R_b / a)^6$ is significantly larger than the detuning $\Delta$, there must exist a single link in the chain such that the sites on both its left and right are unoccupied. For concreteness, we set $N/2$ to be odd  and define a state featuring this ``empty domain wall'' at bond $i$ as $\ket{i}$, where $i = \pm 1, \pm 3, \ldots, \pm N/2$ \footnote{For $N/2$ even, $i = 0, \pm 2, \pm 4 \ldots, \pm N/2$}. These states are referred to as ``hole-type.'' The signs $\mp$ correspond to the left (L) or right (R) sides of the kink, respectively. For illustration, some examples of these states are given below:
\begin{align*}
    \ket{-3} &= \ket{\cdots \CIRCLE\underline{\Circle\Circle} \overset{\text{kink}}{\CIRCLE\Circle|\CIRCLE\Circle} \CIRCLE\Circle \CIRCLE\cdots}, \nonumber\\
    \ket{-1} &= \ket{\cdots \CIRCLE\Circle\CIRCLE \underline{\Circle\Circle} | \CIRCLE\Circle\CIRCLE\Circle \CIRCLE\cdots}, \nonumber\\
    \ket{+1} &= \ket{\cdots \CIRCLE\Circle\CIRCLE \Circle\CIRCLE | \underline{\Circle\Circle} \CIRCLE\Circle \CIRCLE\cdots}, \nonumber\\
    \ket{+3} &= \ket{\cdots \CIRCLE\Circle\CIRCLE \Circle\CIRCLE | \Circle\CIRCLE \underline{\Circle\Circle} \CIRCLE\cdots}.
\end{align*}
Thus, the ground state can be expressed as $|\text{GS}\rangle = \sum_{\text{odd } i} \psi_i \ket{i}$, where $\psi_i = \sin(\frac{\pi i}{N + 4}+\frac{\pi}{2}) / \sqrt{N/2 + 2}$ \cite{liu2024deep}. This sinusoidal distribution arises from the open boundary nature of the array. It indicates a \emph{delocalized} ground state since one can find a domain wall in a $Z$-measurement anywhere in the chain with probability distribution $|\psi_i|^2$.

However, introducing a bond-centered kink modifies the energy landscape of these hole-type configurations, selectively favoring or disfavoring certain states. As $\theta$ decreases, most states with a domain wall far from the center receive an energy penalty of $V_{\text{b},2}(\theta)$ at the kink due to next-nearest neighbor occupation of Rydberg excitations. Conversely, the states labeled by $\ket{\pm 1}$ avoid this energy penalty, making them the lowest energy configurations. Consequently, when probing for domain wall positions in the ground state, there is an increased probability of finding the domain localized at the kink, leading to a highly \emph{localized} profile of the $Z$ operator for the ground state. 

\subsection{Numerical results and effective theory}

\begin{figure}[t]
  \centering
  \includegraphics[width=0.45\textwidth]{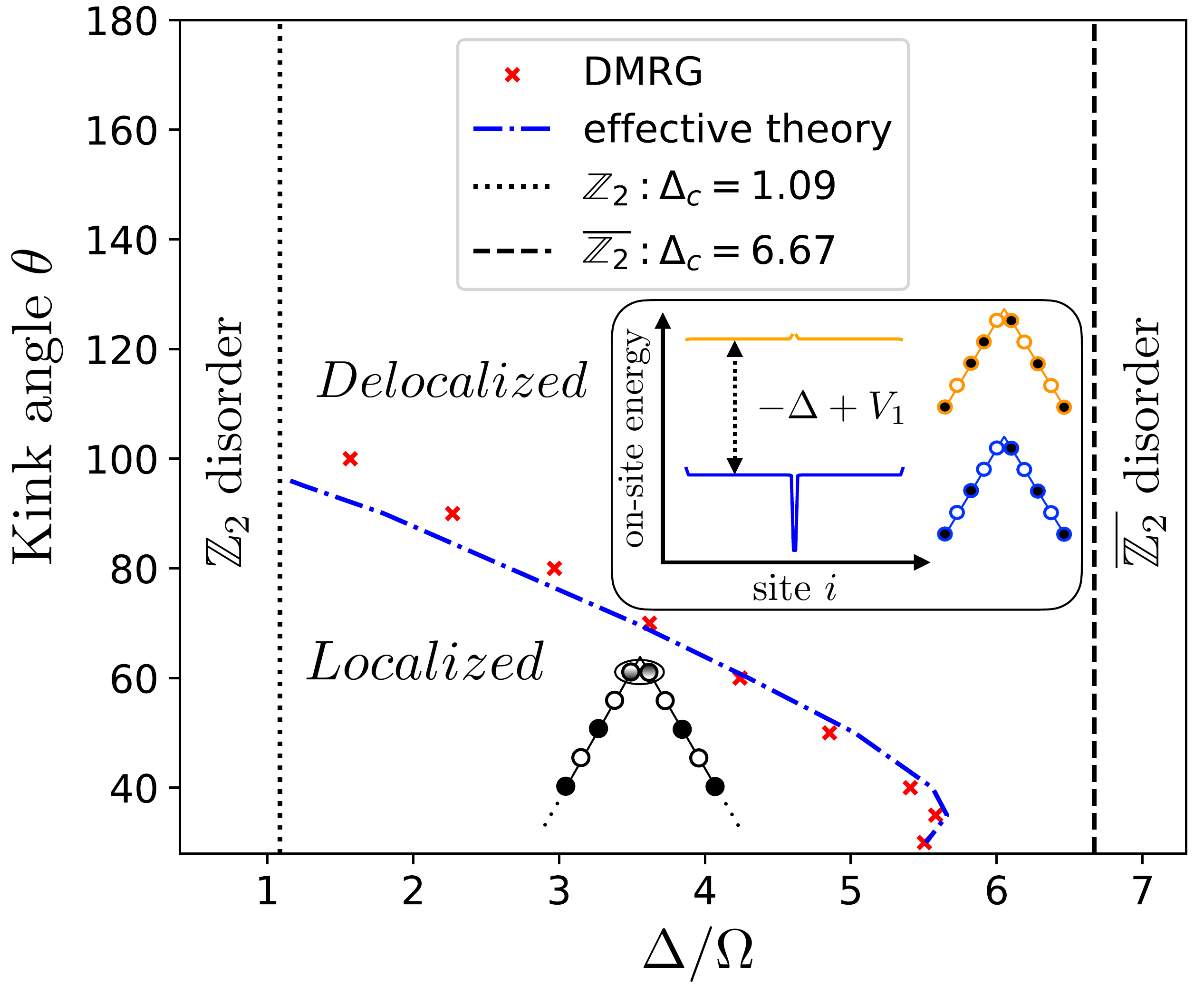}
  \caption{Phase diagram illustrating the localization-delocalization transition. Inset: On-site energy of classical configuration of hole-type (blue) and particle-type (orange).}
  \label{fig:Phase_Diagram}
\end{figure}

We present the phase diagram of this localization-delocalization transition in Fig.~\ref{fig:Phase_Diagram}. The leftmost vertical line refers to the standard $\mathbb{Z}_2$ transition. With large enough detuning, the system will eventually reach the anti-blockade regime of Rydberg atoms, where another $\mathbb{Z}_2$-type transition occurs. One may notice that it is in the same universality class as the first transition, since one can perform the particle-hole transformation $\Circle \leftrightarrow \CIRCLE$; and hence we denote this transition as $\overline{\mathbb{Z}_2}$. At intermediate ranges of detuning, we identify a phase transition boundary between localized and delocalized ground states.

The red crosses in Fig.~\ref{fig:Phase_Diagram} denote transition points obtained from DMRG calculations of the ground state. We measure the transition using the mismatch parameter $\langle \hat{M}_+ \rangle$, where
\begin{align*}
    \hat{M}_+ = -\dfrac{1}{N}\left[\sum_{i=-1}^{-N/2} (-1)^i \hat{Z}_i + \sum_{i=1}^{N/2} (-1)^i \hat{Z}_i\right].
\end{align*}
This operator provides a direct measurement of the delocalization of the domain wall. For example, applying the operator to the state
\begin{align*}
    \cdots \overset{+}{\CIRCLE} \overset{-}{\Circle} \underbrace{\overset{+}{\Circle} \overset{-}{\CIRCLE} \overset{+}{\Circle} \overset{-}{\CIRCLE} \overset{+}{\Circle}}_{\text{mismatch}} \overset{+}{\CIRCLE} \overset{-}{\Circle} \overset{+}{\CIRCLE} \overset{-}{\Circle} \overset{+}{\CIRCLE} \overset{-}{\Circle} \overset{+}{\CIRCLE} \cdots
\end{align*}
enables the detection of mismatches in the sign of $\pm 1$ between the state (filled or empty circles) and the operator ($\pm$ on top), indicative of domain wall locations. When $\langle \hat{M}_+ \rangle = 0$, the domain wall is at the ends of the chain, while for $\langle \hat{M}_+ \rangle \approx 1$, the domain wall is near the kink. Hence, the expectation value $\langle \hat{M}_+ \rangle$ measures the average position of the domain wall. The fluctuation $\langle \hat{M}_+^2 \rangle-\langle \hat{M}_+ \rangle^2$ quantifies its variance, and a larger value of this quantity means a more delocalized state. We expect that this quantity does not scale with system size in the localized phase, while it is proportional to $N^2$ in a delocalized phase.

In numerical simulations (Appendix~\ref{app:wetting}), we indeed observe this scaling behavior when increasing the detuning. Furthermore, we utilize the corresponding Binder ratio and susceptibility to determine the transition point and its scaling exponents. Thus, we conclude there is a second-order phase transition from a localized phase to a delocalized phase. We also examine the fluctuation of domain wall position using the one-point function $\langle\hat{Z}_i\rangle$. It remains finite and of order 1 in the localized phase, while it diverges in the entire delocalized phase. This phenomenon resembles a local effect progressively ``wetting'' the entire chain, either by bending the kink less or increasing the detuning. This wetting transition \cite{huse1982domain,huse1983melting,campostrini2015quantum1,campostrini2015quantum2,hu2021first} represents a pure boundary or defect effect that propagates its influence throughout the entire system.

To understand this localization-delocalization transition, we may model the domain wall as a particle hopping on a chain. For moderate detunings, states containing a single hole-type domain wall form the lowest-energy manifold. However, once the detuning approaches the nearest-neighbor Rydberg blockade strength, states containing a particle-type domain wall—i.e., locally anti-blockaded \cite{ates2007antiblockade,amthor2010evidence,valencia2024rydberg,zhang2024quantum}—come down in energy.
We label these states as $|i\rangle$ at even sites $i=0,\pm2,\cdots$. Examples include:
\begin{align*}
    \ket{-2}=\ket{\cdots \CIRCLE\Circle\underline{\CIRCLE \CIRCLE} \overset{\text{kink}}{\Circle|\CIRCLE}\Circle\CIRCLE\Circle\CIRCLE\cdots}, \nonumber\\
    \ket{0}=\ket{\cdots \CIRCLE\Circle\CIRCLE \Circle\underline{\CIRCLE | \CIRCLE} \Circle\CIRCLE\Circle\CIRCLE\cdots}, \nonumber\\
    \ket{+2}=\ket{\cdots \CIRCLE\Circle\CIRCLE \Circle\CIRCLE | \Circle\underline{\CIRCLE \CIRCLE}\Circle\CIRCLE\cdots}. 
\end{align*}
These particle-type domain wall states have an energy shift of $-\Delta + V_1$ relative to the hole-type ones (see inset of Fig.~\ref{fig:Phase_Diagram}), and the interaction between these two branches, combined with Rabi flips coupling them, drives the transition.

To see that, we construct an effective Hamiltonian 
\begin{equation} \label{eq:two_bands}
  H_\text{eff} =  \sum_i \epsilon_i |i\rangle \langle i| - \frac{\Omega}{2} \sum_{i} \left(|i\rangle\langle i+1|+|i+1\rangle\langle i|\right),
\end{equation}
with on-site energies $\epsilon_i$ given by
\begin{align}
 \epsilon_i = &
  \begin{cases}
  -V_2(\theta)&\, i = \pm1 \\
     \infty&\, i = 0 \\
     0&\, i \in \text{other odd} \\
     -\Delta + V_1&\, i \in \text{other even}\\
  \end{cases}
\end{align}
We begin by integrating out the even sites, which have higher on-site energies, leaving an effective theory on the odd sites. When the gap between the two branches is large, the odd-site energies are only slightly modified, so a potential well remains at $i=\pm1$. In a single-particle picture, any finite well localizes the ground state around that site. As we increase the detuning, the gap $-\Delta + V_1$ shrinks, inducing stronger level repulsion between the two branches. This pushes the lower branch downward at most odd sites—except at $i=\pm1$, which are less affected for two reasons: (i) the gap there is larger, and (ii) site $i=0$, which would otherwise contribute to repulsion, is forbidden. Eventually, the former ``well'' at $i=\pm1$  turns into a ``bump'', allowing the domain wall to delocalize.

This single-particle picture indeed exhibits a localization-delocalization transition in its ground state. Mapping the particle picture back to the domain-wall language and computing the previously discussed mismatch operator $\hat{M}_+$ yields the same critical exponents. The blue dashed line in Fig.~\ref{fig:Phase_Diagram}, indicating the phase boundary, follows from this minimal model and aligns closely with DMRG results, demonstrating that the effective theory captures the key low-energy physics of the transition.

\section{conclusion and outlook}
In this work, we have systematically explored the role of geometric defects in a one-dimensional Rydberg blockade model with $1/R^6$ interactions. By tuning both the system’s detuning parameter and the kink angle, we uncovered a range of nontrivial phenomena. At the critical point, we showed that site-centered kinks effectively cut the chain into two segments, while bond-centered kinks yield distinct RG flows toward intermediate-coupling fixed points. Away from criticality, bond-centered kinks can drive a domain wall localization-delocalization transition, analogous to quantum wetting, thus significantly reshaping the phase diagram. Our findings illustrate that even in a minimal one-dimensional setting, geometric defects can substantially alter the low-energy physics and lead to a variety of intriguing phases and transitions.

Several questions and opportunities arise from these findings. First, extending our analysis to higher-dimensional Rydberg arrays may reveal richer phase diagrams and shed light on the interplay between defects and other boundary effects \cite{kalinowski2022bulk}. Second, investigating the dynamical properties of kinks, potentially via the Kibble-Zurek mechanism \cite{garcia2024quantum}, could offer insights into quantum state preparation and non-equilibrium processes. Finally, the “local probes” introduced here may be used to characterize phases and phase transitions in other strongly correlated systems, both within and beyond Rydberg-based setups.

\begin{acknowledgments}
We thank Wenjun Zhang, Tao Zhang, Yuqing Wang, and Wenlan Chen for helpful discussions on experimental details. We are also grateful to Hui Zhai for his initial collaboration and stimulating ideas. This work is supported by NSFC Grant No.~12342501, and the National Key R\&D Program of China 2023YFA1406702. H.W. is also supported by China Postdoctoral Science Foundation (Grant No.~2024M751609) and Postdoctoral Fellowship Program of CPSF. C.L. is also supported by Tsinghua University Dushi program. Y.G. is also supported by Tsinghua University Dushi program and DAMO Academy Young Fellow program. S.L. acknowledges support from the Gordon and Betty Moore Foundation under Grant No.~GBMF8690, the National Science Foundation under Grant No.~NSF PHY-1748958, and the Simons Foundation under an award to Xie Chen (Award No.~828078). 
The DMRG calculations are performed using the ITensor library \cite{ITensor}.
\end{acknowledgments}

\appendix

\section{\MakeUppercase{Rydberg vs Ising}}
\label{app:Rydberg_vs_Ising}
In this appendix, we perform a comparison between the Rydberg array in PXP limit and the canonical transverse field Ising model. 

We start by exploiting the Ising CFT description of a critical Rydberg array discovered in Refs.~\cite{yao2022quantum} and \cite{slagle2021microscopic}. More precisely, the low-energy physics of a critical Rydberg array is equivalent to that of the critical Ising model described by the following Hamiltonian
\begin{align}
  H_{\rm Ising}=-J\sum_i(\hat{\tau}^z_i\hat{\tau}^z_{i+1}+\hat{\tau}^x_i), 
\end{align}
where $\hat{\tau}^z$ and $\hat{\tau}^x$ refers to the Pauli matrices in the critical Ising with $J>0$. The symmetries of these two models match in an interesting way: (1) Translation by one lattice spacing in Rydberg array is equivalent to the spin flip $\prod_i\hat{\tau}^x_i$ in the Ising chain. (2) The \emph{site-centered} (instead of bond-centered) inversion in the Rydberg array is mapped to the inversion in the Ising chain \footnote{In the critical Ising chain, site-centered and bond-centered inversions are related by a single-site translation and has no essential difference at long distance. In the case of Rydberg chain, however, these two types of inversions are different. }. Related to the first mapping, a closed critical Rydberg array with even (odd) number of sites is equivalent to a critical Ising chain with the periodic (antiperiodic) boundary condition. The second mapping will be useful in a moment for understanding the physics of an open chain. 

Now suppose we add a \emph{site-centered} kink to a closed critical Rydberg array. Let us first focus on the case of even number of atoms $\simeq$ periodic Ising chain. 
The kink obviously breaks the translation symmetry. Hence, in the equivalent Ising description, this kink is mapped to a local perturbation breaking the Ising $\mathbb{Z}_2$ symmetry (generated by the spin flip). A $\mathbb{Z}_2$-breaking local perturbation can be physically understood as a local pinning field coupled to one Ising spin (plus less relevant perturbations). This is described by the Hamiltonian 
\begin{align}
  H_{\rm Ising}'=H_{\rm Ising}+\eta \hat{\tau}^z_1. 
\end{align}
The pinning field $\eta$ is relevant as a defect perturbation. We thus conjecture that the long-distance physics is equivalent to setting $|\eta|=\infty$, i.e. the spin at this site is fixed. It is not hard to see that a closed Ising chain with a fixed spin is equivalent to an open chain with both boundary spins pinned to the same direction. This latter model can then be analyzed using the well-established tools of boundary CFT. We can similarly study the case of odd number of atoms $\simeq$ antiperiodic Ising chain. The antiperiodic boundary condition simply means flipping the sign of one Ising interaction term in $H_{\rm Ising}$, say $\hat{\tau}^z_k\hat{\tau}^z_{k+1}$. This does not affect the aforementioned tricks: The kink is still mapped to a pinning field which effectively breaks the Ising chain. Now, on this open Ising chain, we can cure the sign of $\hat{\tau}^z_k\hat{\tau}^z_{k+1}$ by a unitary transformation, at the cost of reversing the direction of one boundary spin -- the two fixed boundary spins are now opposite to each other. To summarize, we have discovered the following equivalence at the critical point:  
\begin{itemize}
  \item \textit{closed Rydberg array with a kink and even (odd) atoms}
  $\simeq$ \textit{open Ising chain with fixed and aligned (antialigned) boundary spins}. 
\end{itemize}
This result enables us to make quantitative predictions using boundary CFT. 

We note that an open critical Rydberg chain can also be mapped to an open Ising chain with fixed boundary spins \cite{slagle2021microscopic}. The reason is similar: open boundaries of a Rydberg chain break the translation symmetry, and are thus equivalent to boundary pinning fields in the Ising description. Using the previously mentioned mapping between inversion symmetries, one can further infer the following: an open Rydberg chain with odd (even) atoms corresponds to an open Ising chain with fixed and aligned (antialigned) boundary spins. Combining with the previous result, we conclude that at the critical point, 
\begin{itemize}
  \item \textit{closed Rydberg chain with a kink and even (odd) atoms} $\simeq$ \textit{open Rydberg chain with odd (even) atoms}. 
\end{itemize}
This is the ``kink = cut'' statement made at the beginning of this subsection, and is schematically shown in Fig.\,\ref{fig:kink_cut}.

\begin{figure}[t]
    \centering
    \includegraphics[width=0.45\textwidth]{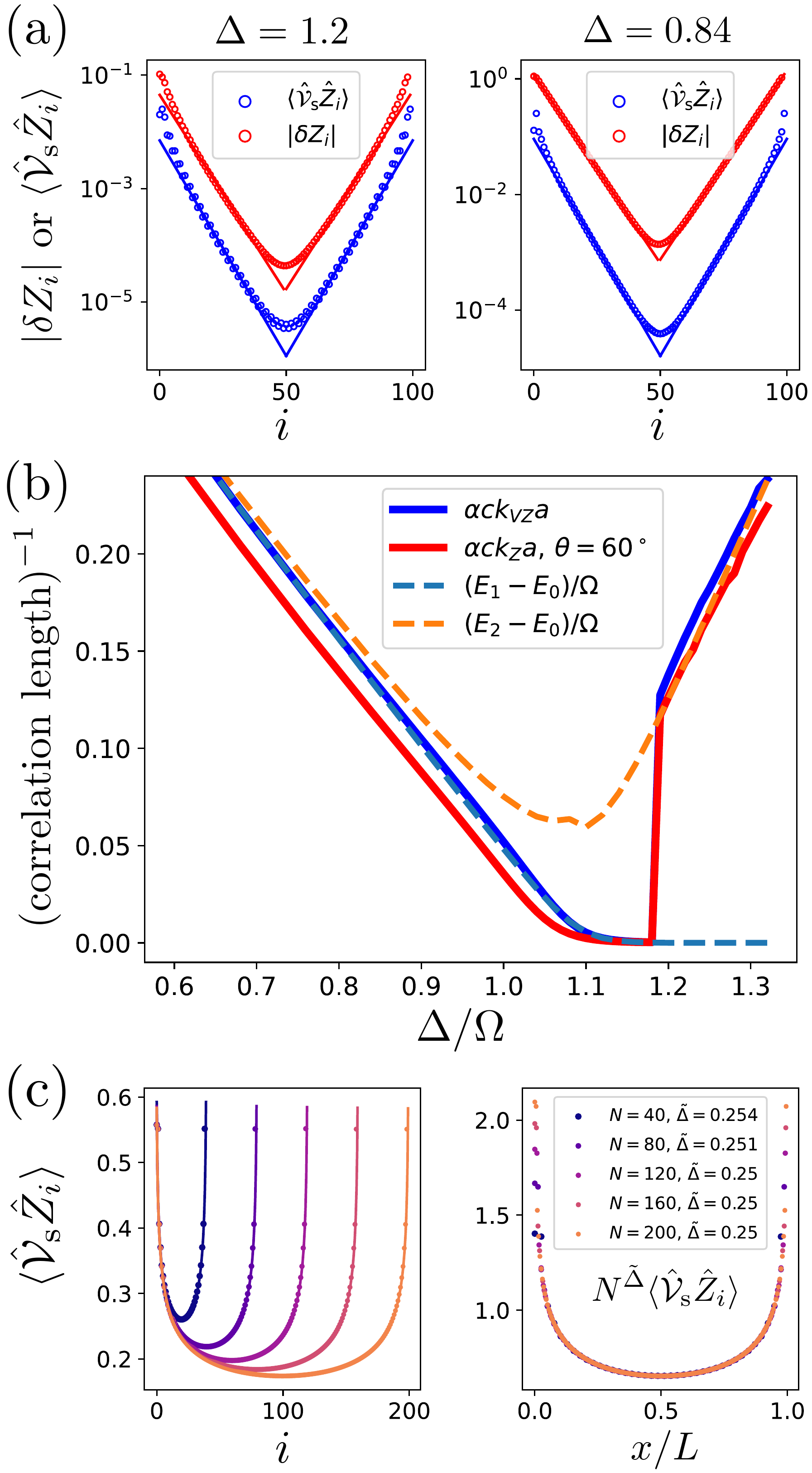}
    \caption{(a) Plot of $|\delta Z_i|$ (in red) and $\langle \hat{\mathcal{V}}_\text{s} \hat{Z}_i \rangle$ (in blue), exhibiting exponential decay with respect to site $i$, for $\Delta > \Delta_c$ (left panel) and $\Delta < \Delta_c$ (right panel). (b) Inverse correlation lengths $k_Z$ for $|\delta Z_i|$ (in red) and $k_{V\!Z}$ for $\langle \hat{\mathcal{V}}_\text{s} \hat{Z}_i \rangle$ (in blue) as functions of detuning $\Delta$. The spectrum of the $1/R^6$ Hamiltonian is included as dashed lines. (c) At the critical point, $\langle \hat{\mathcal{V}}_\text{s} \hat{Z}_i \rangle$ follows a power-law decay with a scaling dimension $\tilde{\Delta} = 1/4$.}
    \label{fig:correlation_length}
\end{figure}

\section{\MakeUppercase{One point correlation with kink vs two point without kink: Perturbation level}}
\label{app:beyond_criticality}
In regimes far from the critical point, the energy gap $E_1-E_0$ becomes substantially large, validating the use of perturbation theory, leading to
\begin{equation}
\label{eq:Z=VZ}
\delta Z_i (\theta) \propto  \langle \text{GS}(\pi)|\hat{\mathcal{V}}_\text{s}\hat{Z}_i|\text{GS}(\pi)\rangle_\text{con}.
\end{equation}
This equation essentially encapsulates the two-point correlation function in a defect-free system within a one-point correlation function featuring a kink. When $\Delta$ deviates significantly from $\Delta_c$, correlations undergo exponential decay, characterized by
\begin{equation}
    \delta Z_i \sim \exp[-k_Z (i a)], \quad \langle \hat{\mathcal{V}}_\text{s} \hat{Z}_i \rangle \sim \exp[-k_{V\!Z} (i a)],
\end{equation}
where $k_Z$ and $k_{V\!Z}$ signify inverse of the correlation length associated with $\delta Z_i$ and $\langle V_k Z_i \rangle$, respectively. The utility of the exponential decay ansatz is corroborated by Fig.~\ref{fig:correlation_length}(a) for both $\Delta > \Delta_c$ (left panel, broken phase) and $\Delta < \Delta_c$ (right panel, unbroken phase). Moreover, the inverse spatial correlation lengths $k_Z$ and $k_{V\!Z}$ are fundamentally related to the energy gap, which serves as the inverse time scale. Considering that space and time are essentially interchangeable up to a speed of light factor $c$, their corresponding correlation lengths should closely resemble one another, i.e., $ck_Za\sim ck_{V\!Z}a\sim \text{(energy gap)}/\Omega$. This is validated by Fig.~\ref{fig:correlation_length}(b), accurate up to an order-1 factor $\alpha$. Even under significant kink effects ($\theta = 60^\circ$), where perturbation theory may break down due to $\langle 1| V_k |0 \rangle \gg E_1 - E_2$, the correspondence expressed by Eq.~(\ref{eq:Z=VZ}) appears to remain largely accurate.

While Eq.~(\ref{eq:Z=VZ}) serves as a useful heuristic for experimental measurements, questions may arise regarding its applicability near the critical point. At criticality, the energy gap diminishes to zero, rendering first-order perturbation theory in a two-level system ineffective. Nonetheless, both quantities do exhibit a power-law decay at the critical point, as substantiated by Fig.~\ref{fig:correlation_length}(c), where we find a scaling dimension $\tilde{\Delta} = 1/4$, contrasting with $\Delta_\phi = 1/8$ for $\delta Z_i$.

\begin{figure}[t]
  \centering
  \includegraphics[width=0.4\textwidth]{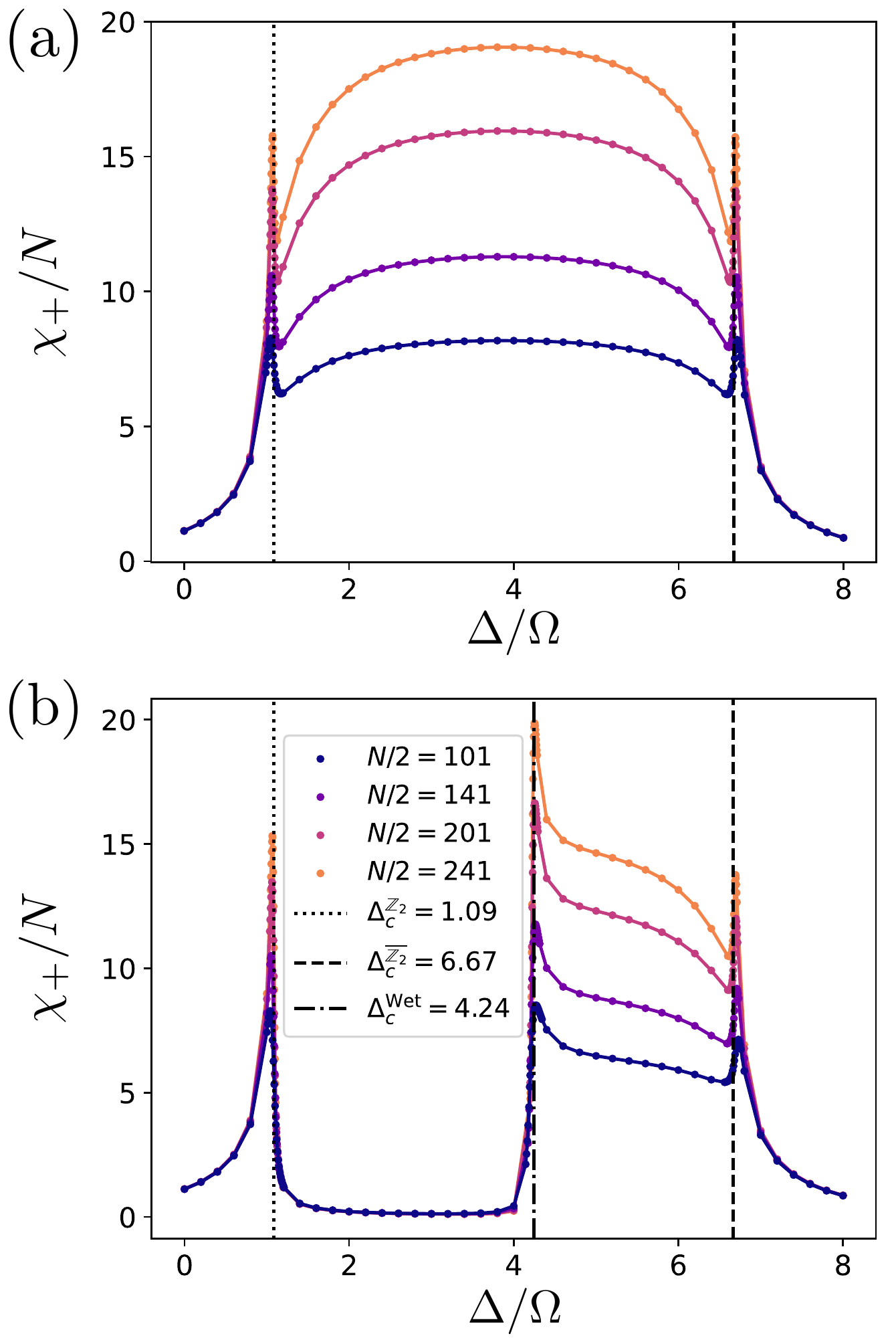}
  \caption{Susceptibility per site $\chi_+/N$ for an even open chain: (a) no kink and (b) a kink angle $\theta=60^\circ$.}
  \label{fig:finite_size_scaling_beyond_critical}
\end{figure}

\begin{figure}[t]
  \centering
  \includegraphics[width=0.5\textwidth]{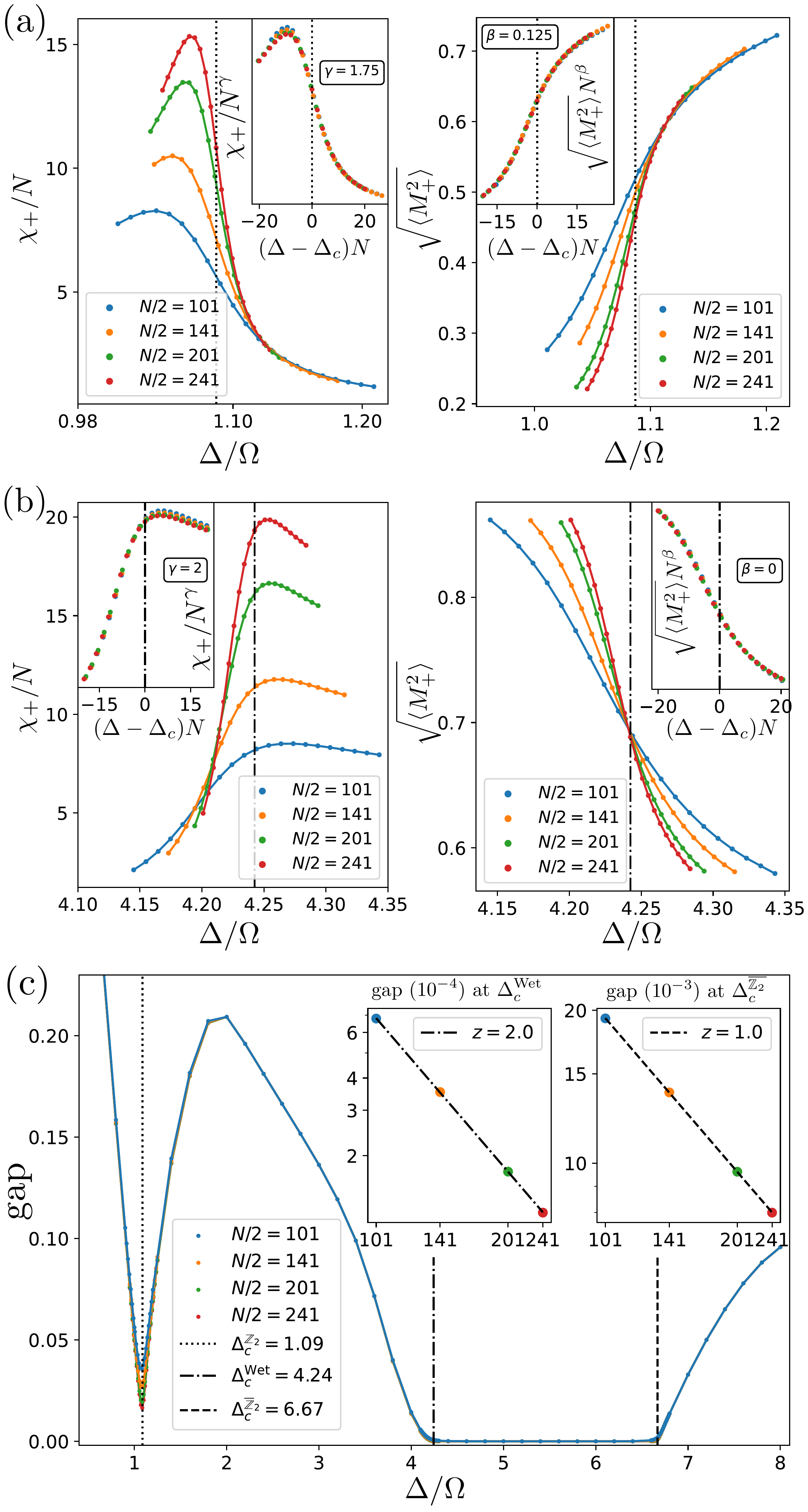}
  \caption{Finite-size scaling of $\chi_+/N$ and $\langle \hat{M}_+^2 \rangle$ in (a) the $\mathbb{Z}_2$ critical regime and (b) the wetting regime. (c) Energy gap vs detuning; inset: log-log plots of the gap vs system size at the wetting and $\mathbb{Z}_2$ transitions.}
  \label{fig:finite_size_scaling}
\end{figure}

\begin{figure}[t]
  \centering
  \includegraphics[width=0.485\textwidth]{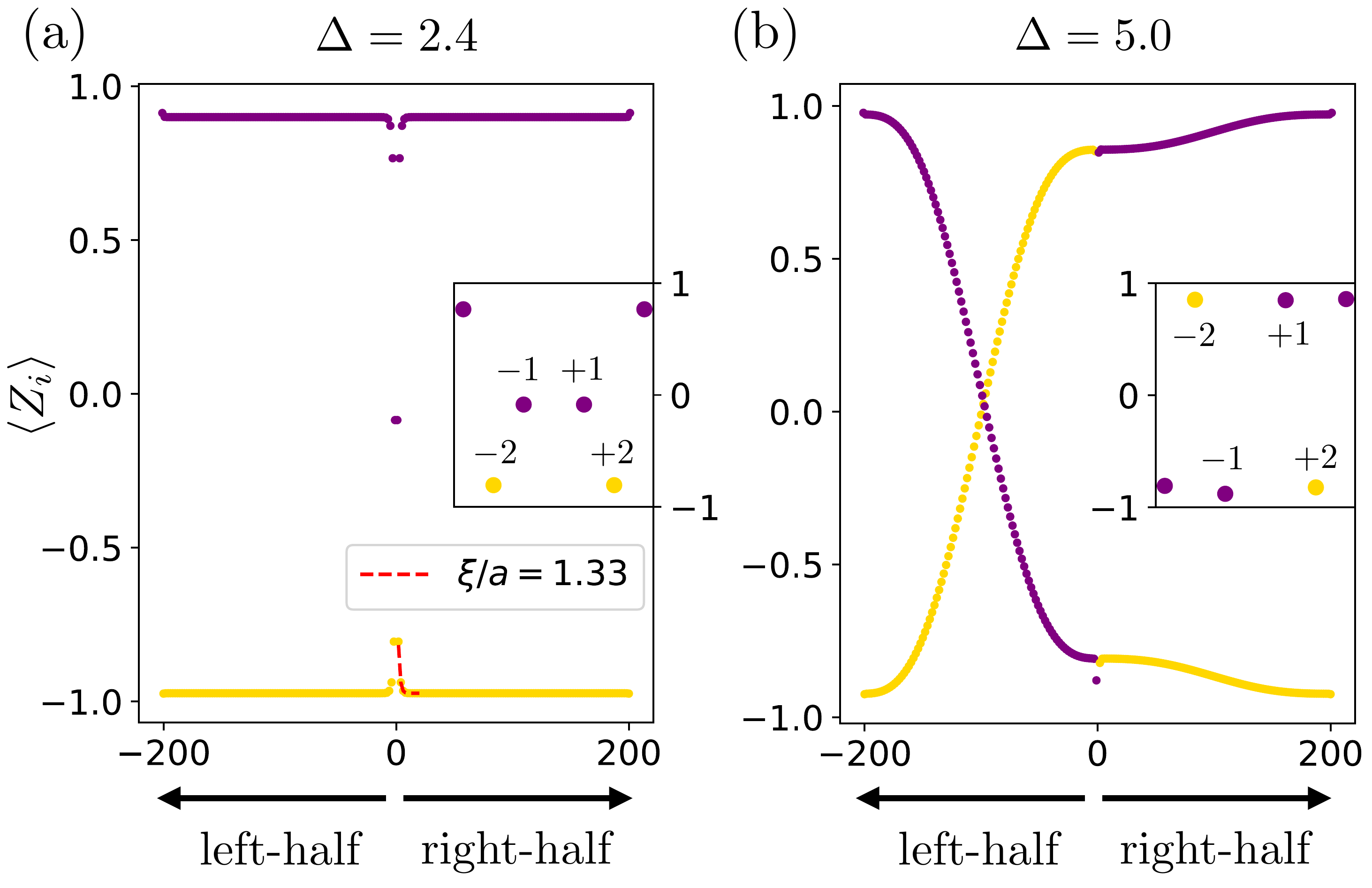}
  \caption{The one-point function $\langle \hat{Z}_i \rangle$ vs site $i$ at $\theta = 60^\circ$. (a) $\Delta < \Delta_c^{\text{Wet}}$ (localized phase) and (b) $\Delta > \Delta_c^{\text{Wet}}$ (delocalized phase). Even (odd) sites are shown in gold (purple).}
  \label{fig:Z_i}
\end{figure}

\section{\MakeUppercase{Numerical Results for Bond-Centered Kinks in open chain}}
\label{app:wetting}

In this appendix, we provide numerical details for a bond-centered kink in an open chain, see Fig.~\ref{fig:kink_cut}(b). 

For $\theta=180^\circ$, Fig.~\ref{fig:finite_size_scaling_beyond_critical}(a) shows how the susceptibility,
$\chi_+ = N^2 \bigl(\langle \hat{M}_+^2 \rangle - \langle \hat{M}_+ \rangle^2\bigr)$,
diverges at the two critical points of the pristine chain. The first one is the usual $\mathbb{Z}_2$ transition at $\Delta = \Delta_c^{\mathbb{Z}_2} = 1.09$. The second occurs at $\Delta = \Delta_c^{\overline{\mathbb{Z}_2}} = 6.67$ and is also of the Ising type, referred to as $\overline{\mathbb{Z}_2}$. This transition can be understood by the transformation $\hat{Z}_i \rightarrow -\hat{Z}_i$ and $\Delta \rightarrow 2(V_1 + V_2) - \Delta$, which leaves the Hamiltonian invariant. A direct estimate based on this symmetry, $
\Delta_c^{\overline{\mathbb{Z}_2}} \approx 2 \times \bigl[(10/8)^6 + (10/16)^6\bigr] - 1.09 = 6.66$, agrees with our numerical value. Figure~\ref{fig:finite_size_scaling}(a) shows the finite-size scaling of $\chi_+ \sim N^{\gamma/\nu}$ and $\langle \hat{M}_+^2 \rangle \sim N^{-2\beta/\nu}$ at $\theta=180^\circ$. We find the 1+1D Ising exponents $\nu=1$, $\beta = 1/8$, and $\gamma = 7/4$.

Figure~\ref{fig:finite_size_scaling_beyond_critical}(b) presents the case of a finite kink angle, $\theta = 60^\circ$. Besides the two Ising critical points, there is now a wetting transition at $\Delta = \Delta_c^{\text{Wet}} = 4.24$. As seen in Fig.~\ref{fig:finite_size_scaling}(b), the critical exponents at this wetting transition are $\nu = 1$, $\beta = 0$, and $\gamma = 2$.

We also extracted the dynamical critical exponent $z$ by examining the energy gap (the difference between the ground-state energy and the first excited state), shown in Fig.~\ref{fig:finite_size_scaling}(c). For the Ising transitions, we find a relativistic value of $z = 1$. In contrast, at the wetting transition, we observe a non-relativistic value of $z = 2$.

For the one-point function $\langle \hat{Z}_i \rangle$, Fig.~\ref{fig:Z_i}(a) shows that for $\Delta < \Delta_c^{\text{Wet}}$, the system is in a localized domain-wall phase. Away from the kink center, $\langle \hat{Z}_i \rangle$ alternates between $\pm 1$, and the domain wall’s position fluctuates over only a few sites ($\xi/a \sim \mathcal{O}(1)$). Zooming in around $i=\pm 1$, one sees cat-like superpositions of the form $\sim |+1\rangle + |-1\rangle$, signaling a “hard” domain wall with a finite excitation gap as shown in Fig.~\ref{fig:finite_size_scaling}(c). 

For $\Delta > \Delta_c^{\text{Wet}}$, Fig.~\ref{fig:Z_i}(b) shows that the domain wall delocalizes over the left half of the chain, while the right half remains in a nearly perfect alternating configuration. This corresponds to a “soft” domain wall phase. The ground state should in principle preserve left-right symmetry, but in practice, DMRG breaks it since the gap between the symmetric and antisymmetric states scales as $1/N^3$, which is too small to be resolved. Hence, we also expect that experimental realizations may favor one side or the other due to inevitable imperfections.

Finally, we briefly comment on how this domain-wall localization transition maps to a quantum particle on a chain with a potential well or bump at the center as discussed in the main text. In the localized phase, the effective potential is a well; in the delocalized phase, it becomes a bump. Exactly at the transition, the potential vanishes, and the free-particle picture suggests the gap scales as $1/N^2$. 
Deep inside the delocalized phase, a classical particle would be confined in the left- or right-half of the chain, with degenerate energy. These two states are connected by tunneling in the quantum case, with an energy splitting $\left[\sin(2\pi/N)/\sqrt{N/2}\right]^2 \sim 1/N^3$.

\bibliography{Rydberg-Kink.bbl}

\end{document}